
%
%
%
%
\input harvmac
%
%
%
%
\ifx\answ\bigans
\else
\output={
  \almostshipout{\leftline{\vbox{\pagebody\makefootline}}}\advancepageno
}
\fi
%
%
%

%
%

%
%
\def\UCSD#1#2{\noindent#1\hfill #2%
\bigskip\supereject\global\hsize=\hsbody%
\footline={\hss\tenrm\folio\hss}}
%
%
\def\abstract#1{\centerline{\bf Abstract}\nobreak\medskip\nobreak\par #1}
%
%
%
%
\edef\tfontsize{ scaled\magstep3}
 \tfontsize  \tfontsize
 \tfontsize \font\titlei=cmmi10 \tfontsize
\font\titleis=cmmi7 \tfontsize \font\titleiss=cmmi5 \tfontsize
\font\titlesy=cmsy10 \tfontsize \font\titlesys=cmsy7 \tfontsize
\font\titlesyss=cmsy5 \tfontsize  \tfontsize
\skewchar\titlei='177 \skewchar\titleis='177 \skewchar\titleiss='177
\skewchar\titlesy='60 \skewchar\titlesys='60 \skewchar\titlesyss='60
%
%
%
%
%
\def\inv{^{\raise.15ex\hbox{${\scriptscriptstyle -}$}\kern-.05em 1}}
\def\lbar{{\lower.35ex\hbox{$\mathchar'26$}\mkern-10mu\lambda}} 

%
%
%
%
\def\dsl{\,\raise.15ex\hbox{/}\mkern-13.5mu D} 
\def\delsl{\raise.15ex\hbox{/}\kern-.57em\partial}
\def\Ksl{\hbox{/\kern-.6000em\rm K}}
\def\Asl{\hbox{/\kern-.6500em \rm A}}
\def\Dsl{\hbox{/\kern-.6000em\rm D}} 
\def\Qsl{\hbox{/\kern-.6000em\rm Q}}
\def\gradsl{\hbox{/\kern-.6500em$\nabla$}}
%
%
\def\lspace{\ifx\answ\bigans{}\else\qquad\fi}
\def\lbspace{\ifx\answ\bigans{}\else\hskip-.2in\fi} 
%
%
\def\boxeqn#1{\vcenter{\vbox{\hrule\hbox{\vrule\kern3pt\vbox{\kern3pt
        \hbox{${\displaystyle #1}$}\kern3pt}\kern3pt\vrule}\hrule}}}
%
%
\def\mbox#1#2{\vcenter{\hrule \hbox{\vrule height#2in
\kern#1in \vrule} \hrule}}
%
%
%
%

  \def\CO{{\cal O}}

%
%
%
%
%

%

\def\bar#1{\overline{#1}}

\def\bra#1{\left\langle #1\right|}
\def\ket#1{\left| #1\right\rangle}

\def\darr#1{\raise1.5ex\hbox{$\leftrightarrow$}\mkern-16.5mu #1}

%
%
\def\frac#1#2{{\textstyle{#1\over #2}}} 
%
%
%
%

\def\Tr{\mathop{\rm Tr}}

%
%
%
%

%
%
\def\ltap{\ \raise.3ex\hbox{$<$\kern-.75em\lower1ex\hbox{$\sim$}}\ }
\def\gtap{\ \raise.3ex\hbox{$>$\kern-.75em\lower1ex\hbox{$\sim$}}\ }
\def\gl{\ \raise.5ex\hbox{$>$}\kern-.8em\lower.5ex\hbox{$<$}\ }
\def\roughly#1{\raise.3ex\hbox{$#1$\kern-.75em\lower1ex\hbox{$\sim$}}}
%
%

%

%

\relax

\noblackbox
\ifx\epsffile\undefined\message{(FIGURES WILL BE IGNORED)}
\def\insertfig#1{}
\else\message{(FIGURES WILL BE INCLUDED)}
\def\insertfig#1{{\centerline{\epsfxsize=4in\epsffile{#1}}}}
\fi
\def\CO{{{\cal O}}}
\def\ln{large $N_c$}
\def\bra#1{\left\langle #1\right|}
\def\ket#1{\left| #1\right\rangle}
\def\nc{N_c}
\def\sref#1{${}^{#1}$}
\rightline{\vbox{\hbox{\tenrm UCSD/PTH 94-01}\hbox{\tenrm hep-ph/9406209}}}
\bigskip
\centerline{{\titlefont{BARYONS IN THE LARGE N LIMIT}}\footnote{*}{Talks
presented at the Symposium on Internal Spin Structure of the Nucleon
(Yale University, Jan 94) and at the Workshop on Continuous Advances in
QCD (Theoretical Physics Institute, University of Minnesota, Feb 94).}
}
\bigskip\bigskip
\centerline{ANEESH V. MANOHAR}
\bigskip
\centerline{\tenit Physics Department, University of California, 9500
Gilman Drive}
\baselineskip=12pt
\centerline{\tenit La Jolla, CA 92093, USA}
\vfill
\abstract{The properties of baryons can be computed in a systematic
expansion in $1/N_c$, where $N_c$ is the number of colors.  Recent
results on the axial couplings and masses of baryons (for the case of
three flavors) are presented. The results give insight into the
structure of flavor $SU(3)$ breaking for baryons.
}
\vfill
\UCSD{UCSD/PTH 94-01}{March 1994}
%
\newsec{Introduction and Summary of Results}

In this talk, I will present some recent results on baryons in the
$1/N_c$ expansion obtained in collaboration with Roger Dashen and
Elizabeth Jenkins.\nref\dm{R.~Dashen and A.V.~Manohar, {\it Phys.\
Lett.\ } {\bf B315} (1993) 425, 438. }\nref\j{E.~Jenkins, {\it Phys.\
Lett.\ } {\bf B315} (1993) 431, 441, 447. }\nref\djm{R.~Dashen,
E.~Jenkins, and A.V.~Manohar, {\it Phys. Rev.\ } {\bf D49} (1994)
4713.}\sref{\dm,\, \j,\, \djm}\ This work shows that
one can use the $1/N_c$ expansion to obtain
predictions for baryon properties that are in good agreement with the
experimental results for $N_c=3$. The results are derived directly from
QCD, and do not make any model assumptions about the structure of baryons.
The success of the predictions indicates that $1/N_c$ corrections are
under control. Other recent work on baryons in the $1/N_c$ expansion can
be found in papers by Carone, Georgi, and Osofsky,\ref\cgo{C.~Carone,
H.~Georgi, and S.~Osofsky, {\it Phys.\ Lett.\ } {\bf B322} (1994) 227.} Luty
and March-Russell,\ref\lmr{M.~Luty and J.~March-Russell, LBL-34778 {\tt
[hep-ph/9310369]}. } and by Broniowski.\ref\b{W.~Broniowski, TPR-93-39
{\tt [hep-ph/9402206]}. } A good introduction to the $1/N_c$ expansion
for QCD are the original papers of 't~Hooft\ref\thooft{G.~'t~Hooft, {\it
Nucl.\ Phys.\ } {\bf B72} (1974) 461, {\it Nucl.\ Phys.\ } {\bf B75}
(1974) 461. }  and Witten,\ref\witten{E.~Witten, {\it Nucl.\ Phys.\ }
{\bf B160} (1979) 57. } and the lecture notes of
Coleman.\ref\coleman{S.~Coleman, {\it Aspects of Symmetry} (Cambridge,
1985).}

The $1/N_c$ expansion can be used to obtain quantitative information
about baryons. The basic approach is to study baryon-meson scattering
amplitudes. One finds that these scattering amplitudes violate unitarity
unless certain cancellation conditions are satisfied. The cancellation
conditions lead to an additional symmetry of baryons in the $N_c=\infty$
limit, which is discussed in the next section. This symmetry allows one
to obtain results for the baryons in terms of the irreducible
representations of the symmetry algebra. The $1/N_c$ corrections can
also be classified in terms of the symmetry algebra. Many of the results
for the case of three light flavors can be obtained without any
assumptions about the mass of the $s$-quark. These results provide
important constraints on the structure of $SU(3)$ breaking in the
baryons. The $1/N_c$ expansion can also be used for baryons containing a
heavy quark.\sref\j\ Some of the results obtained so far
include\sref{\dm,\, \j,\, \djm}
\smallskip
\item{$\bullet$} The baryon sector of QCD has a (contracted) $SU(6)$
spin-flavor symmetry in the large $N_c$ limit. This symmetry can be used
to compute the ratio of pion-baryon couplings, such as
$g_{NN\pi}/g_{N\Delta\pi}$. Similar results hold for the pion couplings
of baryons containing a heavy quark.
\smallskip
\item{$\bullet$} The $F/D$ ratio for the baryon axial currents is
determined to be $2/3 + \CO\left(1/N_c^2\right)$, in good agreement with
the experimental value of $0.58\pm0.04$.
\smallskip
\item{$\bullet$} The $F/D$ ratio for the baryon magnetic moments is determined
to be $2/3 + \CO\left(1/N_c^2\right)$, in good agreement with the experimental
value of $0.72$. The difference between the $F/D$ ratios for the axial
currents and magnetic moments is an indication of the size of $1/N_c^2$
corrections.
\smallskip
\item{$\bullet$} The ratios of all pion-baryon couplings are determined up to
corrections of order $1/N_c^2$, and the ratios of all kaon-baryon couplings are
determined to leading order. These results are independent of the mass of the
$s$-quark.
\smallskip
\item{$\bullet$} The $SU(3)$ breaking in the pion couplings must be linear
in strangeness at order $1/N_c$. This leads to an equal spacing rule for the
pion couplings, which agrees well with the data. The $SU(3)$ breaking in the
decuplet-octet transition axial currents is related to the $SU(3)$ breaking in
the octet axial currents.
\smallskip
\item{$\bullet$} The baryon mass relations
$$\eqalign{
&\Sigma^* - \Sigma = \Xi^* - \Xi \cr
&\frac 1 3 \left( \Sigma + 2 \Sigma^* \right) - \Lambda
= \frac 2 3 \left( \Delta - N \right)\cr
&\frac 3 4 \Lambda + \frac 1 4 \Sigma - \frac 1 2 \left( N + \Xi
\right) = - \frac 1 4 \left( \Omega - \Xi^* - \Sigma^* + \Delta
\right)\cr
&\frac 1 2 \left( \Sigma^* - \Delta \right) - \left( \Xi^* -
\Sigma^* \right) + \frac 1 2 \left( \Omega - \Xi^* \right)=0\cr
&\Sigma^*_Q - \Sigma_Q = \Xi^*_Q - \Xi^\prime_Q\cr
&\frac13\left(2\Sigma^*_Q +\Sigma_Q\right)-\Lambda_Q = \frac23\left(\Delta -
N\right)\cr
}$$
are valid up to corrections of order $1/N_c^2$ without assuming $SU(3)$
symmetry. Some of these relations are also valid using broken $SU(3)$, with
octet symmetry breaking. Relations which can be proved using either large $N_c$
spin-flavor symmetry or broken flavor $SU(3)$ work extremely well, because
effects which violate these relations must break both symmetries.
\smallskip
\item{$\bullet$} The chiral loop correction to the baryon axial
currents cancels to two orders in the $1/N_c$ expansion, and is of order
$1/N_c$  instead of order $N_c$.
\smallskip
\item{$\bullet$} The order $N_c$ non-analytic correction to the baryon masses
is pure $SU(3)$ singlet, and the order one contribution is pure $SU(3)$ octet.
Thus violations of the Gell-Mann--Okubo formula are at most order
$1/N_c$. The baryon masses can be strongly non-linear functions of the
strange quark mass, and still satisfy the Gell-Mann--Okubo formula. This helps
resolve the $\sigma$-term puzzle.
\bigskip
\noindent
I do not have time to discuss all of these results here. I will
concentrate on the pion-baryon couplings and the baryon masses in this
talk. The analysis in Secs.~2 and 3 is for the case of two light flavors
($u$ and $d$). The extension to three flavors is discussed in Sec.~4.

\newsec{Pion-Baryon Couplings}

Baryons in the $1/N_c$ expansion were studied by Witten.\sref\witten\ He
was able to discuss qualitative  features of the baryons, and to show
that (at least for heavy quarks) large $N_c$ baryons could be described
by a Hartree picture. Witten derived $1/N_c$ counting rules for
meson-baryon scattering by studying the $N_c$ dependence of Feynman
diagrams. For example, he showed that the meson-baryon scattering
amplitude is order one, and the baryon-baryon scattering amplitude is
order $N_c$. One can use the qualitative $1/N_c$ counting rules
to obtain consistency conditions on the baryons. These consistency
conditions can be solved to obtain relations for baryons that are
in good agreement with the experimental data at $N_c=3$. The
basic large $N_c$ scaling results I will use are that the pion decay
constant $f_\pi$ is of order $\sqrt{N_c}$, the baryon mass is of order
$N_c$, and the axial coupling of the baryon, $g_A$, is of order $N_c$.
The pion-baryon coupling is of order $g_A/f_\pi\sim\sqrt{N_c}$, if the
interaction is written using a gradient coupling.

\midinsert
\insertfig{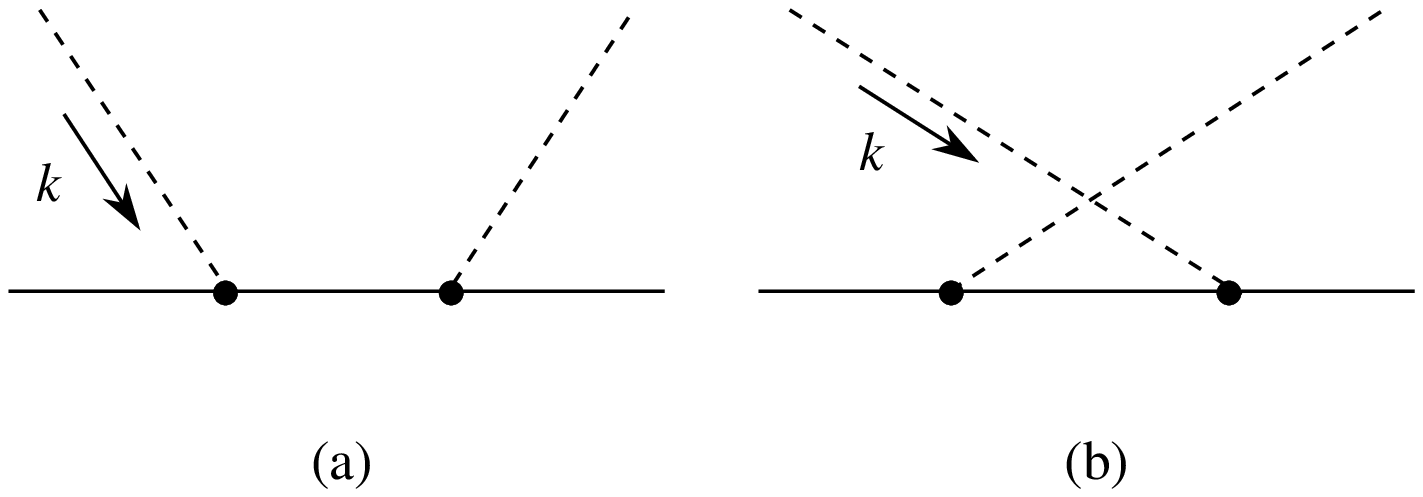}
\vskip-0.75cm
\centerline{\tenrm FIGURE 1.}
\centerline{\tenrm Graphs contributing to pion-baryon scattering at
leading order in $\scriptstyle 1/N_c$.}
\bigskip\bigskip
\endinsert

The leading contributions to pion-baryon scattering in the large $N_c$
limit are given in Fig.~1. The axial current matrix element in the
nucleon can be written as
\eqn\axialmatrix{
\bra{N} \bar\psi\gamma^i\gamma_5 \tau^a\psi \ket{N} = g \nc \bra{N} X^{ia}
\ket{N},
}
where $\bra{N} X^{ia} \ket{N}$ and $g$ are of order one. The coupling
constant $g$ has been factored out so that the normalization of $X^{ia}$
can be chosen conveniently. $X^{ia}$ is an operator (or $4\times4$
matrix) defined on nucleon states $p\uparrow$, $p\downarrow$,
$n\uparrow$, $n\downarrow$, which has a finite \ln\ limit. The
pion-nucleon scattering amplitude for $\pi^a(q) + N (k) \rightarrow
\pi^b(q^\prime)+ N(k^\prime)$ is
\eqn\amp{
-i\ q^i q^{\prime\,j} {\nc^2g^2\over f_\pi^2}
\left[{ 1\over q^0} X^{jb }X^{ia}
- {1 \over q^{\prime\,0} } X^{ia}X^{jb}\right],
}
where the amplitude is written in the form of an operator acting on nucleon
states. Both initial and final nucleons are on-shell, so $q^0=q^{\prime\,0}$.
The product of the $X$'s in Eq.~\amp\ then sums over the possible spins and
isospins of the intermediate nucleon. Since $f_\pi\sim\sqrt\nc$, the overall
amplitude is of order $\nc$, which violates unitarity, and also contradicts the
large $N_c$ counting rules of Witten. Thus \ln\ QCD with a $I=J=1/2$
nucleon multiplet  interacting with a pion is an inconsistent field
theory. There must be other states that cancel the order $\nc$ amplitude
in Eq.~\amp\ so that the total amplitude is order one, and consistent
with unitarity. One can then generalize $X^{ia}$ to be an operator on
this degenerate set of baryon states, with matrix elements equal to the
corresponding axial current matrix elements. With this generalization,
the form of Eq.~\amp\ is unchanged. Thus we obtain the first consistency
condition for baryons,
\eqn\consi{
\left[X^{ia},X^{jb}\right]=0,
}
so that the axial currents are represented by a set of operators
$X^{ia}$ that commute in the \ln\ limit. This consistency condition was
also derived by this method by Gervais and Sakita.\ref\gs{ J.-L.~Gervais
and B.~Sakita, {\it Phys.\ Rev.\ } {\bf D30} (1984) 1795.}\ There are
additional commutation relations,
\eqn\xjcomm{
\left[J^i, X^{jb}\right]=i\,\epsilon_{ijk}\, X^{kb},\qquad
\left[I^a, X^{jb}\right]=i\,\epsilon_{abc}\, X^{jc},
$$
$$
\left[J^i,J^j\right]=i\,\epsilon_{ijk}\, J^k,\qquad
\left[I^a,I^b\right]=i\,\epsilon_{abc}\, I^c,\qquad
\left[I^a,J^i\right]=0,
}
since $X^{ia}$ has spin one and isospin one.

The algebra in Eqs.~\consi\ and \xjcomm\ is a contracted SU(4) algebra.
Consider the embedding $SU(4)\rightarrow SU(2)\otimes SU(2)$ where
$4\rightarrow (2,2)$. If the generators of $SU(2)\otimes SU(2)$ in the
defining representation are $I^a$, and $J^i$, the $SU(4)$ generators in
the defining representation are $J^i\otimes 1$, $1\otimes I^a$ and
$J^i\otimes I^a$, which we will call $I^a$, $J^i$ and $G^{ia}$
respectively. (The properly normalized $SU(4)$ generators are
$I^a/\sqrt{2}$, $J^i/\sqrt{2}$ and $\sqrt2\, G^{ia}$.) The algebra for
\ln\ baryons in QCD is given by taking the limit
\eqn\xlimit{
X^{ia} = \lim_{\nc\rightarrow\infty} {G^{ia} \over \nc},
}
(up to an overall normalization of the $X^{ia}$). The commutation
relations of $SU(4)$,
\eqn\sufour{\eqalign{
&\left[J^i,J^j\right]=i\,\epsilon_{ijk}\,J^k,\cr
&\left[I^a,G^{ia}\right] = i\,\epsilon_{abc}\, G^{jc},\cr
&\left[I^a,J^i\right]=0,}
\qquad\eqalign{
&\left[I^a,I^b\right]=i\,\epsilon_{abc}\, I^c,\cr
&\left[J^i,G^{jb}\right] = i\,\epsilon_{ijk}\, G^{kb},\cr
&\left[G^{ia},G^{jb}\right] = \frac i 4\, \epsilon_{ijk} \delta_{ab}\, J^k +
\frac i 4\,\epsilon_{abc} \delta_{ij}\, I^c,
}}
turn into the commutation relations eqs.~\consi--\xjcomm\ in the \ln\ limit.

The \ln\ limit of QCD has a contracted $SU(4)$ symmetry in the baryon
sector. This explains the success of the non-relativistic $SU(4)$
symmetry of the quark model. The irreducible representations of the
contracted Lie algebra can be  obtained using the theory of induced
representations. The irreducible representation can be labelled by a
quantum number $K=0,\ 1/2,\ 1,\ \ldots$. The $K=0$ irreducible
representation contains baryons with no strange quarks, such as the
nucleon $N$,  the delta $\Delta$, and other baryon states which do not
exist for $N_c=3$. The $K=1/2$ tower has the quantum numbers of baryons
with one strange quark, etc. The contracted Lie algebra is sufficient to
determine the ratio of all the pion-baryon couplings. The normalization
of the couplings is arbitrary, since all the commutation relations are
homogeneous in $X^{ia}$. The explicit form for the pion-baryon couplings
(the matrix elements of $X^{ia}$) can be written using
$6j$-symbols.\sref\djm\

It is easy to show that the \ln\ QCD predictions for the pion-baryon
coupling ratios are the same as those obtained in the non-relativistic
quark model\nref\am{A.V.~Manohar, {\it Nucl.\  Phys.\ } {\bf B248}
(1984) 19.}\nref\karl{G.~Karl and J.E.~Paton, {\it Phys. Rev.\ } {\bf
D30} (1984) 238\semi M.~Cvetic, and J.~Trampetic, {\it Phys.\
Rev.\ } {\bf D33} (1986) 1437.}\sref{\am,\, \karl}\ or in the
Skyrme model,\ref\skyrme{G.S~Adkins, C.R.~Nappi, and E.~Witten, {\it
Nucl.\ Phys.\ } {\bf B228} (1983) 552\semi
N.~Dorey, J.~Hughes, and M.P.~Mattis, {\tt [hep-ph/9404274]}.}\ in the
$N_c\rightarrow \infty$ limit, because both these models also have a
contracted $SU(4)$ symmetry
in this limit. (The contracted $SU(4)$ symmetry is sufficient to
determine the ratios of all the pion-baryon couplings.) In the Skyrme
model, the axial current in the $N_c\rightarrow\infty$ limit is
proportional to $X^{ia}\propto \Tr A \tau^i A^{-1} \tau^a$. The $X$'s
commute, since $A$ is a coordinate, and the expression for $X$ does not
contain any factors of the conjugate momentum. In the quark model, $X$
is proportional to
$$
X^{ia} \propto {1\over N_c} \sum_\alpha q^\dagger_\alpha \sigma^i \tau^a
q_\alpha ,
$$
where the sum on $\alpha$ is over the color index. The commutator of two
$X$'s is
$$\eqalign{
\left[X^{ia},X^{jb}\right]&\propto{1\over N_c^2}\sum_{\alpha\beta}
\left[q^\dagger_\alpha \sigma^i \tau^a
q_\alpha,q^\dagger_\beta \sigma^j \tau^b
q_\beta\right],\cr
&={1\over N_c^2}\sum_{\alpha=\beta}\left[
q^\dagger_\alpha \sigma^i \tau^a
q_\alpha,q^\dagger_\beta \sigma^j \tau^b
q_\beta\right]\rightarrow 0,
}$$
where the last equality follows since quarks of different colors
commute. The sum is at most of order $N_c$, and so vanishes as
$N_c\rightarrow\infty$ because of the overall factor of $1/N_c^2$.

\newsec{$1/N_c$ Corrections}

What makes the $1/N_c$ expansion for baryons interesting is that it is
possible to compute the $1/N_c$ corrections. This allows one to compute
results for the physical case $N_c=3$, rather than for the strict
$N_c=\infty$ limit, which is only of formal interest.

\midinsert
\insertfig{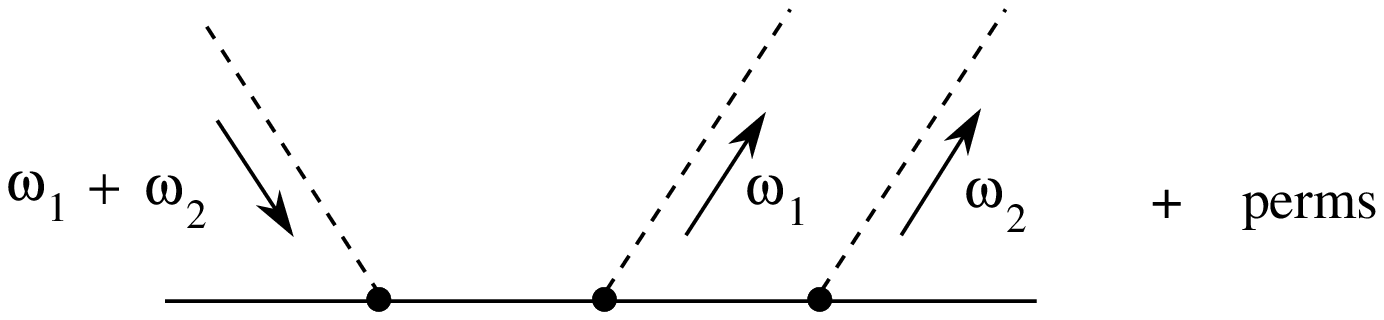}
\vskip-0.75cm
\centerline{\tenrm FIGURE 2.}
\centerline{\tenrm Diagrams contributing to $\scriptstyle \pi +
N\rightarrow \pi+\pi+ N$.}
\bigskip\bigskip
\endinsert

The $1/N_c$ corrections to the axial couplings $X^{ia}$ are determined
by considering the scattering process $\pi^a+N\rightarrow \pi^b+\pi^c+N$
at low energies. The nucleon pole graphs that  contribute in the \ln\
limit are shown in Fig.~2. Define the matrix element of the axial current to
order $1/\nc$ by
\eqn\amatrix{ \bra{N} \bar\psi\gamma^i\gamma_5 \tau^a\psi \ket{N} = g_0 \nc
\bra{N} X^{ia} \ket{N}, }
where $g_0$ is a constant independent of $\nc$. $X^{ia}$ then can be
expanded in a series in $1/\nc$,
\eqn\gexp{
X^{ia}= X^{ia}_0 +{1\over\nc} X^{ia}_1+\ldots
}
(The $X$ used in Sec.~2 is now denoted by$X_0$.) The amplitude for
pion-nucleon scattering from the diagrams in Fig.~2 is proportional to
$$
\nc^{3/2} \left[X^{ia},\left[X^{jb},X^{kc}\right]\right],
$$
and violates unitarity unless the double commutator vanishes at least as
fast as $\nc^{-3/2}$, so that the amplitude is at most of order one. (In fact,
one expects that the double commutator is of order $1/\nc^2$ since the
corrections should only involve integer powers of $1/\nc$. This result also
follows from the \ln\ counting rules which imply that each additional pion has
a factor of $1/\sqrt{\nc}$ in the amplitude.) Substituting eq.~\gexp\ into the
constraint gives
\eqn\acons{
\left[X^{ia}_0,\left[X_1^{jb},X^{kc}_0\right]\right] +
\left[X^{ia}_0,\left[X^{jb}_0,X_1^{kc}\right]\right] = 0, }
using $\left[X^{ia}_0,X^{jb}_0\right]=0$ from Eq.~\consi. The only
solution to the  consistency equation~\acons\ is that $X_1^{ia}$ is
proportional to $X_0^{ia}$. This can be verified by an explicit
computation of $X_1^{ia}$ using reduced matrix elements, or by using
group theoretic methods discussed in ref.~\djm. Thus we find that
\eqn\gans{
X^{ia} = \left(1 + {c\over \nc}\right) X_0^{ia} + \CO\left({1\over
\nc^2}\right),
}
where $c$ is an unknown constant. The first correction to $X^{ia}$ is
proportional to the lowest order value $X_0^{ia}$,  so the $1/\nc$
correction to  the axial coupling constant ratios vanishes. The overall
normalization factor $(1+c/\nc)$ can be reabsorbed into a redefinition
of the unknown axial coupling $g_0$ by the rescaling $g_0 \rightarrow g
= g_0 (1+c/\nc)$, $X^{ia} \rightarrow X_0^{ia}$, so there are no new
parameters at order $1/\nc$ in the axial current sector. The $1/N_c$
rescaling of $g$ depends, in general, on the particular irreducible
representation $K$, i.e. on the strangeness of the baryon. Thus in the
case of three flavors, there is a $1/N_c$ term in the pion-baryon
couplings from $g(K)$, that depends on the strangeness of the baryons.
This $1/N_c$ term can be shown to be linear in $K$, and has a
coefficient which is calculable in the $SU(3)$ limit.\sref\djm

Eq.~\gans\  implies that the commutator $\left[X^{ia},X^{jb}\right]$ is
of order $1/\nc^2$, since there are no $1/N_c$ terms in the expansion of
$X$, and the leading term vanishes by Eq.~\consi. This yields a
pion-baryon scattering amplitude in Eq.~\amp\ of order $1/N_c$, when
one includes the $1/N_c$ corrections for the pion-baryon vertices only.
The full pion-baryon scattering amplitude is expected to be of order
one, not of order $1/N_c$. The order one contribution comes from the
$1/N_c$ correction to the intermediate baryon propagator, that arises
from the baryon mass splittings.\sref\j\ At order $1/N_c$, the baryons
in an irreducible representation of the contracted $SU(4)$ Lie algebra
with a given value of $K$ are no longer degenerate, but are split by an
order $1/N_c$ mass term $\Delta M$. The intermediate baryon propagator
in Eq.~\amp\ should be replaced by $1/(E-\Delta M)$. The energy $E$ of
the pion is order one, whereas $\Delta M$ is of order $1/N_c$, so the
propagator can be expanded to order $1/N_c$ as
\eqn\propexp{
{1\over E-\Delta M} = {1\over E} + {\Delta M\over E^2} +\ldots
}
The order one contribution to the pion-baryon scattering amplitude
arises from the second term in the expansion of the propagator, and is
proportional to $\left[X^{ia},\left[X^{jb},\Delta M\right]\right]$.
Including the $1/N_c$ corrections to the propagator does not affect the
derivation of Eq.~\consi, as the two terms in Eq.~\propexp\ have
different energy dependences. The first term leads to the consistency
condition Eq.~\consi\ and the second gives the consistency condition on
the baryon masses,\sref{\j,\, \djm}
\eqn\mcons{
\left[X^{ia},\left[X^{jb},\left[X^{kc},\Delta M\right]\right]\right]=0.
}
This constraint can be used to obtain the $1/N_c$ corrections to the
baryon masses. The constraint Eq.~\mcons\ is equivalent to a simpler
constraint obtained by Jenkins using chiral perturbation theory\sref\j
\eqn\mconsII{
\left[X^{ia},\left[X^{ia},\Delta M\right]\right]={\rm constant}.
}
The solution of Eq.~\mcons\ or \mconsII\ is that the baryon mass
splitting $\Delta M$ must be proportional to $J^2/N_c=j(j+1)/N_c$, where
$j$ is the  spin of the baryon.

\newsec{The Extension to Three Flavors}

The analysis so far has concentrated on the two flavor case. The $1/N_c$
expansion for baryons for three flavors is more complicated than for two
flavors, because the flavor $SU(3)$ representation of the baryons changes
with the number of colors. The $SU(3)$ weight diagram of the spin-1/2
baryons for $N_c$ colors is shown in Fig.~3. For $N_c=3$, this reduces
to the familiar weight diagram for the baryon octet.

\midinsert
\insertfig{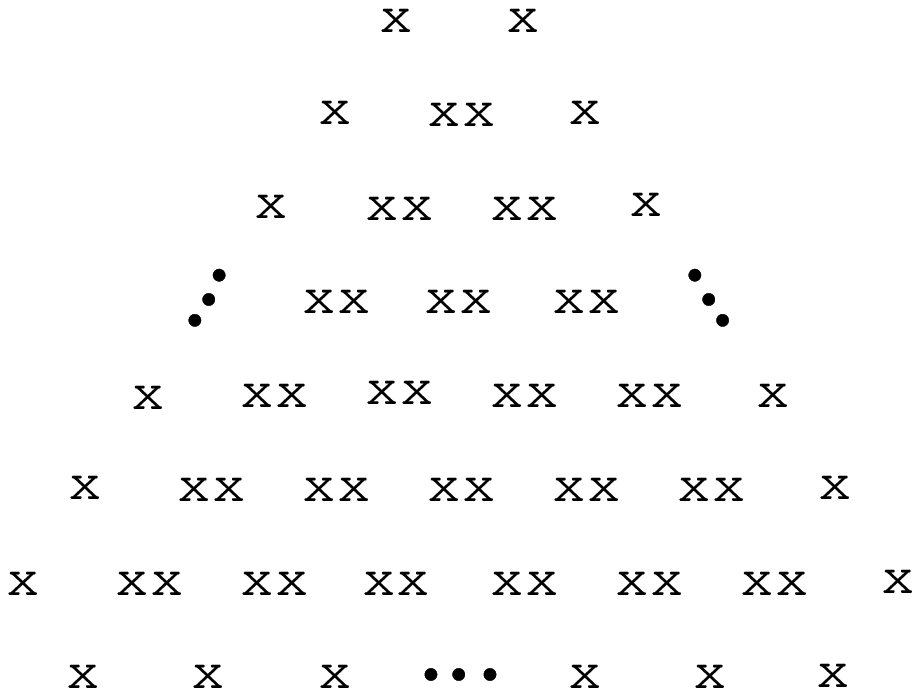}
\vskip-0.75cm
\centerline{\tenrm FIGURE 3.}
\centerline{\tenrm The $\scriptstyle SU(3)$ weight diagram for the
spin-1/2 baryons for $\scriptstyle N_c$}
\vskip -2pt
\centerline{\tenrm colors. The long edge of the triangle has $\scriptstyle
N_c-1$ states.}
\bigskip\bigskip
\endinsert

The method used in Ref.~\djm\ for three flavors was to work with the
$SU(2)$ of isospin, and use the results of the previous sections for the
different strangeness (i.e. $K$) sectors. The different strangeness
sectors can then be related using $K$-meson--baryon scattering, in
essentially the same manner as the analysis of Jenkins\sref\j\ for
baryons containing heavy quarks. The disadvantage of this method is that
it does not use a manifestly $SU(3)$ invariant formalism. However, the
big advantage is that this method can be used without assuming $SU(3)$
symmetry, which allows one to analyze the structure of $SU(3)$ breaking
in the baryons in the $1/N_c$ expansion. The constraints on $SU(3)$
breaking from the $1/N_c$ expansion are extremely interesting, and help
resolve some long-standing puzzles about why $SU(3)$ symmetry works so
well in the baryon sector. We will see that for some quantities, $SU(3)$
breaking is suppressed by powers of $1/N_c$ or $1/N_c^2$.

\subsec{$F/D$}

The $K=1/2$ baryon tower contains the $S=-1$ baryons, and includes two
spin-1/2 baryons, the $\Lambda$ and $\Sigma$. The ratio of the pion
couplings within a given tower is known to order $1/N_c^2$. The explicit
results from Ref.~\djm\ give
\eqn\fdi{
{\Sigma^+\rightarrow \Sigma^0 \pi^+\over \Sigma^+\rightarrow \Lambda^0
\pi^+} = 1 + \CO\left({1\over N_c^2}\right)
}
This result can be compared with the result for $N_c=3$ in the $SU(3)$
limit, $\sqrt 3F/D$, to give $F/D=1/\sqrt{3}=0.58$, which is in
excellent agreement with the experimental value.  One can show that the
$F/D$ ratio of $2/3$ obtained using the non-relativistic quark model is
also consistent with all the pion-baryon couplings up to corrections of
order $1/N_c^2$. Thus one could equally well use $2/3=0.67$ for the
predicted value of $F/D$. There is an ambiguity in the value of $F/D$ at
order $1/N_c^2$, which is connected with the fact that \ln\ flavor
representation of the baryons depends on $N_c$ for three or more
flavors. We will use the quark model value of $2/3$ in what follows.
This value is in good agreement with the experimental $F/D$ ratio of
$0.58\pm0.04$.\ref\jaffe{R.L.~Jaffe and A.V.~Manohar, {\it Nucl.\ Phys. } {\bf
B337} (1990) 509.}

The $1/N_c$ results for the baryon magnetic moments are very similar to
those for the pion-baryon couplings. The $F/D$ ratio is again predicted
to be $2/3$ plus corrections of order $1/N_c^2$, and is in good
agreement with the experimental value of 0.72. The difference between
the experimental values of $F/D$ for the baryon magnetic moments and
pion couplings is an indication of the size of $1/N_c^2$ effects in
QCD---they appear to be small, so that the $1/N_c$ expansion in the
baryon sector appears to be under control.

\subsec{Masses}

The baryon mass constraints can be analyzed for three flavors. Only the
results will be presented here. It can be shown that the baryon masses
must have the form\sref{\j,\, \djm}
\eqn\bform{
M = N_c\ a + b\ K + {1\over N_c}\ \left(c\ I^2 + d\ J^2 + e\ K^2\right)
+\CO\left({1\over N_c^2}\right),
}
where $a$--$e$ are constants which have an expansion in powers of
$1/N_c$. This equation is valid without assuming $SU(3)$ symmetry, and
holds for arbitrary values of the $s$-quark mass. It provides some
interesting information on $SU(3)$ breaking in the baryon masses.

Eq.~\bform\ leads to the baryon mass relations\sref\djm
\eqn\mrelns{\eqalign{
&\Sigma^*-\Sigma=\Xi^*-\Xi + \CO\left({1\over N_c^2}\right),\cr
\noalign{\smallskip}
&{3\Lambda+\Sigma\over 4}-{N+\Xi\over2}=-{1\ \over4}\left[\left(\Omega
- \Xi^*\right) -\left(\Sigma^*-\Delta\right)\right]+ \CO\left({1\over
N_c^2}\right),\cr
\noalign{\smallskip}
&{1\over2}\left(\Sigma^*-\Delta\right)+{1\over2}\left(\Omega-\Xi^*\right)
=\Xi^*-\Sigma^* + \CO\left({1\over N_c^2}\right),
}}
which are valid in the $1/N_c$ expansion irrespective of the value of
$m_s$. One can also derive mass relations using $SU(3)$ perturbation
theory in $m_s$, without using the $1/N_c$ expansion. This leads to the
well-known relations valid including terms of first order in $m_s$,
\eqn\mrelnsii{\eqalign{
&{3\Lambda+\Sigma\over 4}-{N+\Xi\over2}={\rm GMO}=0,\cr
\noalign{\smallskip}
&\left(\Omega - \Xi^*\right) -\left(\Sigma^*-\Delta\right)={\rm ESRI} =0,\cr
\noalign{\smallskip}
&{1\over2}\left(\Sigma^*-\Delta\right)+{1\over2}\left(\Omega-\Xi^*\right)
-\Xi^*-\Sigma^* = {\rm ESRII}=0,
}}
which are the Gell-Mann--Okubo formula, and the two equal spacing rules.
Comparing Eqs.~\mrelns\ with Eqs.~\mrelnsii\ we see that ESRII is true
in either the $1/N_c$ or the $m_s$ expansions. It is only violated at
higher order than $m_s^2/N_c^2$, and so works extremely well
(LHS=146~MeV, RHS=149~MeV).\foot{In principle, the $m_s$ expansion has
non-analytic terms of order $m_s^{3/2}$. This contribution vanishes for
ESRII.\ref\jmasses{E. Jenkins, {\it Nucl.\ Phys.\ } {\bf B368} (1992)
190.}}\ Similarly, the difference between GMO and ESRI is only violated
at order $m_s^{3/2}/N_c^2$, and also works extremely well (LHS=7~MeV,
RHS=3~MeV). The violation of ESRII and ${\rm GMO}-{\rm ESRI}$, which is
of second order in both $m_s$ and $1/N_c$, is comparable in size to
isospin violating effects.  The GMO relation is respected by the order
$N_c$ and order one terms, and is violated in the $1/N_c$ expansion at
order $1/N_c$ rather than $1/N_c^2$. It works about as well as the
difference between GMO and ESRI. This means that a particular linear
combination of the coefficients in Eq.~\bform\ is about a third, instead
of being one. The $1/N_c$ expansion partly explains the success of this
relation; the other part is an accidental cancellation. The relation
$\Sigma^*-\Sigma=\Xi^*-\Xi$ is valid only in the $1/N_c$ expansion, but
not in the $m_s$ expansion, and does not work as well (LHS=191~MeV,
RHS=215~MeV) as the relations that are valid in both expansions. A
systematic study of the mass relations obtained at different orders in
the $1/N_c$ and $m_s$ expansions shows that $SU(3)$ breaking effects are
of order 25\%, and $1/N_c$ effects are of order $1/3$. Some relations,
such as the GMO relation, work slightly better than this naive estimate.
However, none of the relations has an  unexpectedly large $1/N_c$
correction, which is a signal for the breakdown of the $1/N_c$ expansion.

\newsec{Conclusions}

The $1/N_c$ expansion provides a systematic method of computing the
properties of the baryons. The results are in good agreement with the
experimental values at $N_c=3$ when the $1/N_c$ corrections are
included. Whether this success holds for other quantities remains to be
seen. The $1/N_c$ expansion also helps explain the nature of $SU(3)$
breaking in the baryons. There are some interesting results on
non-linear $SU(3)$ breaking effects, and the connection with chiral
perturbation theory,\sref\djm\ which I do not have time to discuss here.
At present, the $1/N_c$ expansion can be used to find the operator form
of $1/N_c$ corrections, but not the absolute normalization.

The connection between $SU(3)$ symmetry and the $1/N_c$ corrections is
subtle, and needs to be explored further. Other approaches to the
$1/N_c$ expansion for baryons,\sref{\cgo,\, \lmr} than the one presented
here give additional insight. An analysis of the Adler-Weisberger sum
rule in \ln\ can be found in Ref.~\b. Hadron scattering in the
$N_c\rightarrow \infty$ limit has been studied in
\nref\mattis{M.P.~Mattis and M. Mukerjee, {\it Phys. Rev. Lett.\ }
{\bf 61} (1988) 1344\semi M.P.~Mattis and E.~Braaten, {\it Phys. Rev.\
}{\bf D39} (1989) 2737.} Ref.~\mattis, which derives an $I_t=J_t$ selection
rule for meson-baryon scattering.

\newsec{Acknowledgements}
The work presented in this talk was done in collaboration with R.~Dashen
and E.~Jenkins. This work was supported in part by the Department of
Energy under grant number DOE-FG03-90ER40546 and by the National Science
Foundation under PYI award PHY-8958081.

\listrefs
\bye